\documentclass[10pt,twocolumn,prl,byrevtex,showpacs,bibnotes]{revtex4}
\usepackage{amsfonts}

\usepackage{graphics}

\begin{document}

\setlength{\parskip}{0pt}%

\title{Comment on ``First-order phase transitions: equivalence between bimodalities
\\and the Yang-Lee theorem''}%

\author{Hugo Touchette}
%

\email{htouchet@alum.mit.edu}%

\address{\mbox{School of Mathematical Sciences, Queen Mary, University of London,
London E1 4NS, UK}}%

\date{\today}%

\begin{abstract}%

I discuss the validity of a result put forward recently by Chomaz and
Gulminelli [Physica A 330 (2003) 451] concerning the equivalence of two
definitions of first-order phase transitions. I show that distributions of
zeros of the partition function fulfilling the conditions of the Yang-Lee
Theorem are \textit{not necessarily} associated with nonconcave
microcanonical entropy functions or, equivalently, with canonical
distributions of the mean energy having a bimodal shape, as claimed by
Chomaz and Gulminelli. In fact, such distributions of zeros can also be
associated with concave entropy functions and unimodal canonical
distributions having affine parts. A simple example is worked out in detail
to illustrate this subtlety.

\end{abstract}%


\maketitle%

Chomaz and Gulminelli \cite{chomaz2003} have studied recently the
equivalence of two different definitions of first-order phase
transitions---one based on the nonconcavity of the microcanonical entropy
function and another based on the distribution of the zeros of the partition
function. My goal here is to question and correct one of their results with
the help of a counterexample which I will then explain using some basic
results of convex analysis. To start, let me summarize the main results
found in Ref.%
~%
\cite{chomaz2003}. The notation I will be using throughout is less general
than the one used in \cite{chomaz2003}; it is simpler, but captures
nevertheless the essence of the problem.

Consider an $n$-body system with energy $U$ and mean energy $u=U/n$. The
partition function of the system is defined as 
\begin{equation}
Z_n(\beta )=\int \Omega _n(u)e^{-\beta nu}du,
\end{equation}
where $\Omega _n(u)$ represents the density of microstates with mean energy $%
u$. It is well-known from the work of Yang and Lee \cite{yang1952} that one
way to make sense of nonanalytic points of the canonical free energy
function 
\begin{equation}
\varphi (\beta )=\lim_{n\rightarrow \infty }-\frac 1n\ln Z_n(\beta ),
\end{equation}
which signal the onset of phase transitions, is to study the distribution of
complex zeros of $Z_n(\beta )$ in the limit $n\rightarrow \infty $. In the
case of first-order phase transitions, in particular, it is known that, as $%
n\rightarrow \infty $, there is an accumulation of zeros of $Z_n(\beta )$
around some positive real value $\beta _c$ of the inverse temperature
corresponding to the value at which $\varphi (\beta )$ is nondifferentiable,
and that the loci of zeros in the vicinity of $\beta _c$ is parallel to the
imaginary axis. This phenomenon is what Chomaz and Gulminelli refer to as
the Yang-Lee Theorem. I shall refer to it myself as the \textit{Yang-Lee
Condition} (YLC). Thus we say that a first-order phase transition appears in
the thermodynamic limit of the canonical ensemble when the zeros of $%
Z_n(\beta )$ satisfy YLC.

Now, what Chomaz and Gulminelli purported to show in \cite{chomaz2003} is
that YLC\ is equivalent to another definition of first-order phase
transitions based on the bimodal shape of the density function $\Omega _n(u)$
(see \cite{chomaz2003} and references therein). They showed first that if $%
\Omega _n(u)$ has a bimodal shape which persists as $n\rightarrow \infty $,
a condition which I shall refer to as the \textit{Bimodal Condition} (BC),
then YLC is satisfied. Actually, the quantity which they focused on was not $%
\Omega _n(u)$ but the canonical distribution $p_{n,\beta }(u)$, given by 
\begin{equation}
p_{n,\beta }(u)=\frac{\Omega _n(u)e^{-\beta nu}}{Z_n(\beta )}\approx \frac{%
e^{-n[\beta u-s(u)]}}{Z_n(\beta )}.
\end{equation}
Yet since a bimodality of $\Omega _n(u)$ translates into a bimodality of $%
p_{n,\beta }(u)$, and vice versa, it does not matter which quantity we refer
to, and, for the purpose of the presentation, I shall stick to $\Omega _n(u)$%
.

It should be noted in passing that the result ``BC implies YLC'' had already
been proved by Lee \cite{lee1996}. The real novelty of \cite{chomaz2003} is
to attempt to prove the converse result, namely that YLC\ implies BC. It is
this second result that Chomaz and Gulminelli claim to have proven, but
which cannot be true in fact, as exemplified by the following counterexample.

\begin{figure*}
\centering
\resizebox{0.8\textwidth}{!}{\includegraphics{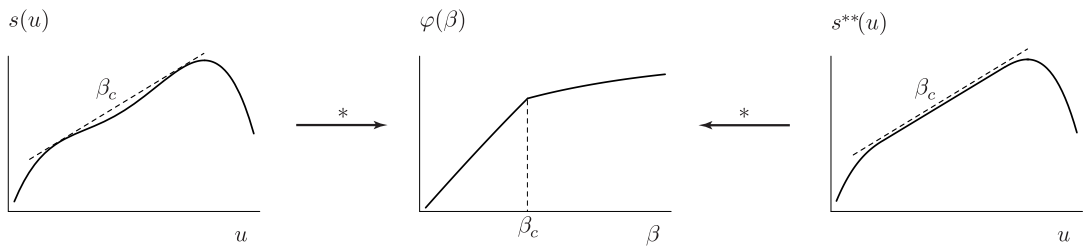}}
\caption{A generic nonconcave entropy function $s(u)$ and its concave
envelope $s^{**}(u)$, which is affine (with slope $\protect\beta_c$) where $s(u)$ is
nonconcave. Both entropies have the same Legendre-Fenchel transform which
corresponds to the free energy function $\protect\varphi (\protect\beta )$.}
\label{entfig}
\end{figure*}%

Consider a density of microstates of the form 
\begin{equation}
\Omega _n(u)=\left\{ 
\begin{array}{lll}
e^{n\Delta } &  & u\in [0,\Delta ] \\ 
0 &  & \text{elsewhere,}
\end{array}
\right.
\end{equation}
with $\Delta >0$. The partition function for this form of $\Omega _n(u)$ is
trivially evaluated and has for expression: 
\begin{equation}
Z_n(\beta )=\frac{e^{n\Delta }}n\left( \frac{1-e^{-\beta n\Delta }}\beta
\right) .
\end{equation}
Setting $Z_n(\beta )=0$ and solving for $\beta \in \Bbb{C}$, we find that
the zeros of the partition function must solve the equation $e^{-\beta
n\Delta }=1$ with the exclusion of $\beta =0$. This is equivalent to $%
e^{-\beta n\Delta }=e^{\pm 2\pi ik}$, $k=1,2,\ldots $, and so we find the
zeros of $Z_n(\beta )$ to be given by $\beta _k=\pm 2\pi ik/(n\Delta )$, $%
k=1,2,\ldots $. In terms of the fugacity $z=e^{-\beta }$, these can be
re-expressed as 
\begin{equation}
z_k=\exp \left( \pm \frac{2\pi ik}{n\Delta }\right) ,\qquad k=1,2,\ldots .
\end{equation}

Now comes the contradiction: $\Omega _n(u)$ does not satisfy BC, but the
zeros of the partition do satisfy YLC. In fact, the zeros of $Z_n(\beta )$
are all aligned on the imaginary axis and pinch the real axis at the
critical value $\beta _c=0$ as $n\rightarrow \infty $. Following the
Yang-Lee theory \cite{yang1952}, we then know that $Z_n(\beta )$ must
develop a nonanalytic point at $\beta _c$ as $n\rightarrow \infty $, which
translates into a nondifferentiable point of $\varphi (\beta )$. This can be
verified by a direct calculation of the free energy: 
\begin{eqnarray}
\varphi (\beta ) &=&\lim_{n\rightarrow \infty }-\frac 1n\ln \left[ \frac{%
e^{n\Delta }}n\left( \frac{1-e^{-\beta n\Delta }}\beta \right) \right] 
\nonumber \\
&=&\left\{ 
\begin{array}{lll}
-\Delta &  & \beta >0 \\ 
-\Delta +\beta \Delta &  & \beta \leq 0.
\end{array}
\right.
\end{eqnarray}
The left- and right-derivatives of the free energy at $\beta _c=0$ being
equal to $\varphi ^{\prime }(\beta =0^{-})=\Delta $, $\varphi ^{\prime
}(\beta =0^{+})=0$, respectively, we also find that the latent heat for this
example is equal to $\Delta $.

The important conclusion that we reach from this simple counterexample is
clear: YLC does not imply BC in general, as claimed in \cite{chomaz2003}. To
determine what the correct implication should be, I shall recall at this
point three important results of equilibrium statistical mechanics and
convex analysis (see \cite{ellis2000,touchette2004}):

(i) If $\Omega _n(u)$ has a bimodal shape that persists when $n\rightarrow
\infty $, then $s(u)$ must be nonconcave over some range of mean energy. The
converse statement is also true.

(ii) The free energy function $\varphi (\beta )$ is always the
Legendre-Fenchel transform of the microcanonical entropy function $s(u)$; in
symbols, 
\begin{equation}
\varphi (\beta )=\inf_u\{\beta u-s(u)\}.
\end{equation}
This holds no matter what the shape of $s(u)$ is, be it concave or not.

(iii) Regions of mean energies over which $s(u)$ is nonconcave or is affine
(i.e., is a line) are indicated at the level of the canonical free energy $%
\varphi (\beta )$ by the existence of points of $\varphi (\beta )$ where
this function is nondifferentiable. The fact that affine parts of $s(u)$
also translate into nondifferentiable points of $\varphi (\beta )$ can be
understood by noting that $s(u)$ and its concave envelope $s^{**}(u)$,
defined by 
\begin{equation}
s^{**}(u)=\inf_\beta \{\beta u-\varphi (\beta )\},
\end{equation}
have the same Legendre-Fenchel transform, namely, 
\begin{equation}
\varphi (\beta )=\inf_u\{\beta u-s(u)\}=\inf_u\{\beta u-s^{**}(u)\};
\end{equation}
see Fig.%
~%
\ref{entfig}.

These results indicate altogether that a first-order phase transition in the
canonical ensemble emerges from the point of view of the microcanonical
ensemble in basically two ways: either $s(u)$ is nonconcave somewhere or
else $s(u)$ is affine somewhere \cite{touchette2004}. The first possibility
is what was considered in \cite{chomaz2003}, whereas the second possibility
is what I have considered in the counterexample, and what is precisely
absent from \cite{chomaz2003}. Hence at this point we have all the
ingredients to conjecture what the correct relationship between YLC and BC
is, namely: if the zeros of $Z_n(\beta )$ satisfy LYC, then either $\Omega
_n(u)$ is bimodal or else it has an affine part. Let me refer to the second
possibility as the \textit{Affine Part Condition} (APC). Then, the result is
simply: YLC\ implies BC\ or APC. Combining this result with what we already
had, namely that BC implies YLC, we arrive finally at the following: \textit{%
YLC is satisfied if and only if BC or APC\ is satisfied}. (This result
should be understood as a conjecture rather than a rigorously proved result
because, at this point, it remains to rigorously prove that YLC is a
necessary and sufficient condition for $\varphi (\beta )$ to have a
nondifferentiable point.)

In the end, one may wonder whether entropies having affine parts are
anything to worry about. There seems indeed to be a lack of structural
stability inherent with such entropies, and so it is questionable whether
they can show up in realistic models. This concern, however, is out of the
scope of the present comment. Affine entropies present one with a
theoretical possibility that one has to take into account when deriving
general theoretical results.

\begin{acknowledgments}%

This work was supported by NSERC (Canada) and the Royal Society of London.

\end{acknowledgments}%

\end{document}